\documentclass[journal=acs,manuscript=article]{achemso}
\usepackage[utf8]{inputenc}
\usepackage{graphicx}
\usepackage{caption}
\usepackage{subcaption}
\usepackage{mhchem}
\usepackage{array}
\usepackage{tabularx}
\usepackage{float}
\usepackage{xcolor}
\usepackage[numbers,sort&compress]{natbib}
\usepackage{mciteplus}
\usepackage{pdfpages}

\title{Highly Antioxidative Lithium Salt Enables High-Voltage Ether Electrolyte for Lithium Metal Battery}

\author{Jian He}
\affiliation{Materials Chemistry and Catalysis, Debye Institute for Nanomaterials Science, Utrecht University, 3584 CG Utrecht, The Netherlands}
\author{Shihan Qi}
\affiliation{School of Physics and Electronics, Hunan University, China}
\author{Jianmin Ma}
\affiliation{School of Physics and Electronics, Hunan University, China}
\email{nanoelechem@hnu.edu.cn}
\author{Nongnuch Artrith}
\affiliation{Materials Chemistry and Catalysis, Debye Institute for Nanomaterials Science, Utrecht University, 3584 CG Utrecht, The Netherlands}
\email{n.artrith@uu.nl}

\begin{document}

\maketitle

\section*{Abstract}

\hspace*{1em} Ether-based electrolytes exhibit excellent compatibility with Li metal anodes, but their instability at high voltages limits their use in high-voltage Li metal batteries. To address this issue, we introduce an alternative perfluorobutane sulfonate (LiPFBS)/dimethoxyethane (DME) electrolyte to stabilize DME in a 4.6~V Li$||$LCO battery. Our study focuses on the formation of solid-electrolyte interphase (SEI) and cathode-electrolyte interphase (CEI) layers compared to the LiTFSI/DME electrolyte. We demonstrate that LiPFBS helps maintain DME’s compatibility in SEI formation. Additionally, a durable CEI layer derived from \ce{PFBS^{-}} enhances the performance of the cell at high voltages by forming a robust, inorganic-dominant CEI layer. A \ce{PFBS^{-}}-derived CEI significantly enhances the overall performance of the full cell under high voltage conditions.

\section*{Introduction}

\hspace*{1em} Electric vehicles and portable devices demand energy storage systems with higher energy densities. Current lithium-ion batteries are nearing their theoretical capacity density limits, prompting the shift towards Li metal batteries\cite{janekSolidFutureBattery2016,linRevivingLithiumMetal2017,albertusStatusChallengesEnabling2018, fanDualInsuranceDesign2019, chenOpportunitiesChallengesHighenergy2020, jiangProbingFunctionalityLiFSI2022,mpupuniRevolutionizingLithiumMetal2024,halldin_stenlid_computational_2024}. However, the practical application of Li metal anode is challenged by the instability of the solid-electrolyte interphase (SEI) caused by conventional carbonate electrolytes. The electrochemically unstable SEI can crack easily during cycling, leading to 'dead Li' and capacity loss\cite{chenDeadLithiumMass2017}. The ideal SEI/CEI should exhibit stability under high voltage to ensure both high energy density and stability simultaneously.

Ether-based electrolytes have been found to be highly compatible with Li metal anodes in early research due to their decent stability against Li metal\cite{kochStabilitySecondaryLithium1978,foosLithiumCyclingPolymethoxymethane1983,peledElectrochemicalBehaviorAlkali1979}. Recent efforts have been dedicated to improving the cycle life in Li metal batteries, including the use of electrolyte additives \cite{jiaEnablingEtherBasedElectrolytes2020} and fluoridation \cite{yuRationalSolventMolecule2022,MolecularDesignElectrolyte,zuInsightLithiumMetal2015} to further promote DME application in Li metal batteries. However, DME is unstable at high voltage because it cannot form a stable cathode-electrolyte interphase (CEI) layer, thereby hindering its application in high voltage cathode systems \cite{SelfheatingInducedHealing, jiaoStableCyclingHighvoltage2018,zhangFundamentalUnderstandingLithium2021,miaoNewEtherbasedElectrolyte2016}.  The high cut-off voltage can lead to irreversible, continuous electrolyte side reactions, resulting in rapid capacity decay \cite{zhangDynamicEvolutionCathode2018, baiFormationLiFrichCathodeelectrolyte2022, yangArmorlikeInorganicrichCathode2023, wangCatalyticallyInducedRobust2023}. To suppress these continuous side reactions, engineering of the solid electrolyte interphase (SEI) and cathode-electrolyte interphase (CEI) is crucial. An ideal SEI/CEI should be an electronic insulator, a fast Li-ion conductor, and mechanically robust to hinder dendrite growth on the anode and volume change in the cathode. However, solvent-derived CEIs are generally unstable\cite{yuRationalSolventMolecule2022}, leading to continuous electrolyte oxidation. Highly concentrated electrolytes were initially introduced to prevent Al corrosion\cite{mcowenConcentratedElectrolytesDecrypting2014} in the 3.5–4 V (vs \( \mathrm{Li}^+ \)/Li) range. LiTFSI has shown its potential in high voltage Li batteries\cite{yuInsightEffectsCurrent2022,liDualsaltsLiTFSILiODFB2016,dinhDeepEutecticSolvent2020,abu-lebdehHighVoltageElectrolytesBased2008}. Meanwhile, recent studies show that high concentration ether-based electrolytes have good oxidation stability at high voltage\cite{renHighconcentrationEtherElectrolytes2019,pengSolventMoleculeReconstruction2022,liangEtherbasedElectrolyteChemistry2021,jiaoStableCyclingHighvoltage2018}. Our recent work using perfluorinated sulfonates as an electrolyte additive showed good compatibility in Li metal batteries\cite{qiElectrolytesEnrichedPotassium2021}. Perfluorinated sulfonates can participate in CEI formation and stabilize LiNi$_{0.6}$Mn$_{0.2}$Co$_{0.2}$O$_2$ (NMC622) in a voltage range of 3-4.5V. Therefore, with the intention of stabilizing DME at high voltage, we designed a lithium perfluorobutane sulfonate (LiPFBS)/dimethoxyethane (DME) electrolytes for 4.6V Li$||$LCO full cell, and comprehensively investigated the mechanism of SEI/CEI formation.

\section*{Method}

\subsection*{Classical molecular dynamics simulation}

All classical molecular dynamics (MD) simulations were executed with the Large Scale Atomic/Molecular Massively Parallel Simulator (LAMMPS)\cite{thompsonLAMMPSaFlexibleSimulation2022}. For all electrolyte systems, electrolytes were first packed randomly in a cubic box with periodic boundaries of size 80 × 80 × 80 Å$^3$ using PACKMOL\cite{martinezPACKMOLPackageBuilding2009} and allowed to optimize in an NPT ensemble simulation at 300~K. Subsequently, an NVT ensemble simulation at 300~K was conducted for 10~ns. The deafult Nosé-Hoover thermostat and barostat were used. A convergence criterion of \(1.0 \times  10^{-4}\)~eV was adopted to minimize the initial configuration, and the sampling of the radial distribution functions was carried out in an NVT ensemble at 300~K for another 10~ns.
The parameters for non-bonded and bonded interactions (bonds, angles, dihedrals, and improper dihedrals) across all structures were derived from the Optimized Potentials for Liquid Simulations All Atom (OPLS-AA) force fields\cite{jorgensenDevelopmentTestingOPLS1996}. Atomic partial charges were calculated by fitting the molecular electrostatic potential at atomic centres in Gaussian16\cite{frischGaussian16Revision2016} using the HF/6-31G(d) level of theory\cite{DevelopmentOPLSAAForce}.

\subsection*{Density Functional Theory (DFT)}

For DFT electronic structure calculations of molecules in \textbf{Figure~\ref{fig:1}}, all molecule geometries were optimized using the Gaussian16 package\cite{frischGaussian16Revision2016}, the M062X exchange-correlation functional\cite{zhaoM06SuiteDensity2008}, and a 6-31G(d) basis set\cite{hehreSelfConsistentMolecular1972}. Single-point energies were obtained using the M062X functional and the def2tzvp basis set\cite{weigendBalancedBasisSets2005} with the universal solvation model based on solute electron density (SMD)\cite{marenichUniversalSolvationModel2009}. The convergence criteria of \texttt{SCF=tight} were used in all Gaussian calculations.
For reduction potential calculations, the electrode reduction potential E was defined via the Nernst equation as
\begin{equation}
E = -\frac{\Delta G^{\mathrm{red}}}{n F} - 1.4 \, \mathrm{V}
\end{equation}
where $F$ is the Faraday constant, $n$ is the number of electrons transferred in the redox reaction, and 1.4~V is the potential offset corresponding to the lithium metal reference electrode. This offset accounts for the difference between the lithium metal reference and the standard hydrogen electrode (SHE).
The free energy of reduction under implicit solvation model $\Delta G_{\mathrm{sol}}^{\mathrm{red}}(\mathrm{O})$ was calculated as
\begin{equation}
\Delta G_{\mathrm{sol}}^{\mathrm{red}}(\mathrm{O}) = \Delta G_{\mathrm{gas}}^{\mathrm{EA}} + \Delta G^{\text{solv}}\left(\mathrm{R}^{-}\right) - \Delta G^{\text{solv}}(\mathrm{O})
\end{equation}
where $\Delta G_{\mathrm{gas}}^{\mathrm{EA}}$ represents the electron affinity in the gas phase, $\mathrm{R}^{-}$ represents the reduced species and $\mathrm{O}$ represents the neutral species. For the solvation free energy $\Delta G^{\text{solv}}$, all molecules were optimized at the B3LYP\cite{leeDevelopmentColleSalvettiCorrelationenergy1988}/6-311++G(d,p) level of theory because the B3LYP functional can provide a good balance between computational cost and accuracy of geometries and frequencies\cite{GeometriesVibrationalFrequencies}. Single-point energies were calculated at the M062X/def2TZVP level of theory for the gas phase and at the M052X/6-31G(d) level of theory for the solvation energy. A previous comprehensive benchmark study by Marenich et al.\cite{marenichUniversalSolvationModel2009} found M052X/6-31G(d) to give accurate solvation energies.

To investigate the electrolyte-\ce{LiCoO2} interface, the $10\overline{1}4$ surface was selected because it is the most stable non-polar surface\cite{nohComparisonStructuralElectrochemical2013d, UnderstandingOnsetSurface} and allows Li ion de-/intercalation.
All calculations for the \ce{LiCoO2} $10\ensuremath{\overline{1}}4$ surface with adsorbed \ce{TFSI^{-}}/\ce{PFBS^{-}} models and \ce{TFSI^{-}}/\ce{PFBS^{-}} in vacuum shown in \textbf{Figure~\ref{fig:6}a} were performed using the Fritz Haber Institute ab initio materials simulations package (FHI-aims)\cite{blumInitioMolecularSimulations2009} with the r$^{2}$SCAN functional\cite{furnessAccurateNumericallyEfficient2020d}.
A $1\times{}1\times{}1$ k-point mesh with the predefined \emph{light} basis set (4th order expansion of the Hartree potential, radial integration grids with 302 points in the outer shell, and a tier 1 basis set) were used.
Dispersive interactions were accounted for with the Tkatchenko-Scheffler van der Waals correction\cite{tkatchenkoAccurateMolecularVan2009}.
The scalar relativistic atomic ZORA correction was employed for all calculations.

To determine low-energy configurations of \ce{TFSI^{-}} and \ce{PFBS^{-}} on the \ce{LiCoO2} $10\ensuremath{\overline{1}}4$ surface at 0~K, \ce{TFSI^{-}}/\ce{PFBS^{-}} was randomly distributed on the \ce{LiCoO2} surface, and the topmost surface layer of \ce{LiCoO2} and the \ce{TFSI^{-}} and \ce{PFBS^{-}} molecules were allowed to relax. The adsorption energy was calculated as
\begin{equation}
\Delta E_{\text{ads}}
= E_{\text{adsorbate/slab}} - E_{\text{slab}} - E_{\text{adsorbate}}
  \label{eq:Ead}
\end{equation}
where $E_{\text{adsorbate/slab}}$ is the DFT energy of the surface slab with adsorbate and $E_{\text{slab}}$ and $E_{\text{adsorbate}}$ are the energies of the bare surface and the isolated adsorbate molecule.

The Fermi energies of each structure were used as the zero point of the energy for the analysis of orbital energies, so that the highest occupied molecular orbital (HOMO) and the lowest unoccupied molecular orbital (LUMO) of the isolated and adsorbed  \ce{PFBS^{-}}/\ce{TFSI^{-}} molecules were defined as
\begin{equation}
E_{\text{HOMO/LUMO, relative}} = E_{\text{HOMO/LUMO, absolute}} - E_{\text{Fermi}}
\end{equation}
where $E_{\text{HOMO/LUMO, relative}}$ is the relative HOMO/LUMO energy, $E_{\text{HOMO/LUMO, absolute}}$ is the absolute HOMO/LUMO energy of isolated \ce{PFBS^{-}}/\ce{TFSI^{-}} molecules, and $E_{\text{Fermi}}$ is the Fermi energy of each structure, respectively. For \ce{PFBS^{-}}/\ce{TFSI^{-}} adsorption on the \ce{LiCoO2}($10\ensuremath{\overline{1}}4$) surface, the projected density of states (PDOS) was used for comparison with the local HOMO/LUMO, as shown in Figure~S6.

\subsection*{Preparation of Electrolytes}

The lithium salts, LiPFBS and LiTFSI, were purchased from Tokyo Chemical Industry Co., Ltd. (Shanghai), and the solvent, DME, was bought from Sinopharm Chemical Reagent Co., Ltd. Different amounts of lithium salts were added to the DME solvent to prepare electrolytes with varying concentrations. All materials were used without further purification or processing. When a larger amount of lithium salt was required, the addition process was performed slowly, ensuring that the previously added lithium salt was fully dissolved before adding more. If the salt dissolved slowly, the system temperature could be raised, but it should not exceed 45 °C. All preparation processes were conducted in a glove box (\ce{O2} $<$ 0.1 ppm, \ce{H2O} $<$  0.1 ppm). The final high-concentration electrolyte was a transparent, viscous, homogeneous liquid.

\subsection*{Electrochemical Measurements}

Li$||$Li symmetric cells were assembled using two Li foils (diameter 16 mm, purchased from China Energy Lithium Co., Ltd.) as the working and counter electrodes in a 2025-type coin cell. The \ce{LiCoO2} (LCO) powder was obtained from Shanshan New Energy Technology Co., Ltd. To prepare the LCO electrode, LCO, carbon black, and polyvinylidene fluoride (PVDF) were mixed in N-methyl-2-pyrrolidone (NMP, analytical reagent) to prepare a slurry. This slurry was then coated on Al foil and dried in a vacuum oven at 80 °C for 24 hours. The dried foil was punched into disks with a diameter of 12 mm for use as LCO cathodes. The capacity was calculated based on the total mass of the entire electrode disk. The separator used was commercial Celgard 2325 (Celgard, LLC). All cell assembly processes were conducted in a glove box (\ce{O2} $<$ 0.1 ppm, \ce{H2O} $<$ 0.1 ppm). Li$||$LCO full cells were assembled using Li foil and LCO electrodes.

Constant current tests for cell performance were conducted using a Neware Battery Testing System (CT-4008, Shenzhen, China) at room temperature (25 °C). Li$||$Li symmetric cells underwent galvanostatic charge and discharge tests at a current density of 0.5 mA cm\(^{-2}\) with a capacity of 0.5 mAh cm\(^{-2}\), o measure lithium deposition and stripping capabilities. The growth of polarization voltage indicated lithium dendrite layer formation. Li$||$LCO cells were galvanostatically tested within a voltage range of 3 to 4.5 V (vs. \ce{Li^{+}}/Li). Cycling performance was tested at a rate of 1C, and the rate capability was tested at 0.2C, 0.5C, 1C, and 2C. An IVIUM electrochemical workstation was used to measure electrochemical impedance spectroscopy (EIS) and cyclic voltammetry (CV). EIS testing employed a two-electrode method, combining the reference electrode and counter electrode into one. In symmetric cells, each lithium metal electrode could function as the working, counter, or reference electrode. In Li$||$LCO cells, the lithium metal negative electrode acted as both the counter and reference electrode, while the LCO electrode served as the working electrode. The test frequency range was 106 Hz to 10\(^{-2}\) Hz. Cyclic voltammetry was used to study reactions occurring within the cells. The sweep voltage ranged from 3 to 4.5 V (vs. \ce{Li^{+}}/Li) with a sweep rate of 2 mV s\(^{-1}\).

\subsection*{Characterization}

The morphology of LCO particles and the lithium metal anode after cycling was observed using a scanning electron microscope (SEM). The SEM model used was the Hitachi S4800, operating at a voltage of 5 kV. Both the cathode and anode samples were cleaned with DME prior to observation to prevent residual electrolyte from affecting the results. The samples were cut into 5 mm by 5 mm pieces and attached to a metal stage for observation. During transfer, both the cathode and anode samples were placed in a container filled with Ar gas to minimize exposure to air, avoiding the impact of oxygen and moisture. SEM analysis revealed surface film formation at the interfaces of both the cathode and anode.

X-ray photoelectron spectroscopy (XPS) tests were conducted using the MDTC-EQ-M20-01 instrument. All XPS spectra were referenced to adventitious carbon at 248.8 eV. CasaXPS\cite{fairleySystematicCollaborativeApproach2021} was used to analyze the XPS spectra. Similarly, during transportation, the samples were placed in a container filled with argon gas to minimize contact with air. The analysis of chemical compositions on the electrode surface was conducted in situ, with different Ar sputtering times of 0, 2, and 4 minutes.

\section*{Results \& Discussion}

Interface issues dominate the overall performance of Li metal batteries. Building upon prior investigations into the influence of solvation structure on electronic properties\cite{heUnveilingRoleLi2021,zhangElectrolyteSolvationStructure2021,tianElectrolyteSolvationStructure2022,liSolvationStructureTuning2023}, it is believed that SEI f5ormation is governed by \ce{Li^{+}} solvation. The electrochemical behavior and the redox potentials, therefore, cannot be understood by only analyzing the HOMO and LUMO of isolated molecules. \ce{Li^{+}} solvation not only affects the molecular orbitals but also the potential energy surface of reactive species on the electrode surface. Therefore, to investigate the reductive behavior of a liquid-state electrolyte, accounting for \ce{Li^{+}} solvation is essential.

\textbf{Figure~\ref{fig:1}} shows the HOMO and LUMO of DME, \ce{PFBS^{-}} and bis(tri\-fluoro\-methane\-sulfonyl)imide (\ce{TFSI^{-}}) with varying degrees of \ce{Li^+} coordination, as obtained from quantum-chemical calculations. Specifically, isolated, singly Li-coordinated, and doubly Li-coordinated configurations were examined for DME, \ce{PFBS^{-}}, and \ce{TFSI^{-}}. An increase in the degree of coordination leads to analogous reductions in both the HOMO and the LUMO for DME and \ce{PFBS^{-}}.

A lower LUMO energy indicates a preference for reduction at the anode, and the electronic structure of the electrolyte components is, therefore, a critical factor in the formation of the SEI. Conversely, a higher HOMO energy indicates preferential oxidation at the cathode, leading to the formation of the CEI layer. Lithium-ion coordination significantly lowers the LUMO energy levels of species in the electrolyte, indicating that species coordinated with lithium are more likely to participate preferentially in the formation of the SEI in a LiPFBS/DME electrolyte system.

In this work, DME as a solvent was expected to be reduced before any anion species because of its lower LUMO energy in different coordination environments. However, as seen in \textbf{Figure \ref{fig:1}}, the range of LUMO energies of both \ce{TFSI^{-}} and \ce{PFBS^{-}} overlap with that of DME, indicating that these two species also have the potential to participate in SEI formation.

\begin{center}
\includegraphics[width=1.06\textwidth]{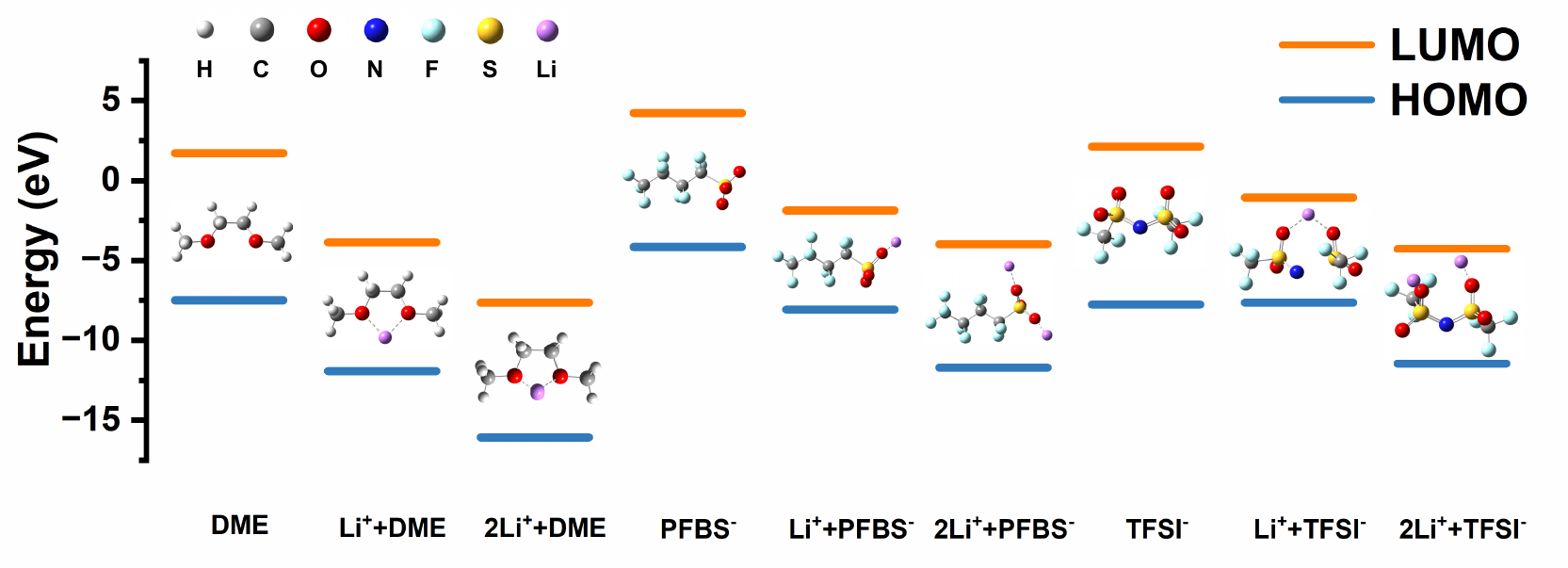}
\captionof{figure}{LUMO and HOMO energy for DME, \ce{TFSI^{-}}, and \ce{PFBS^{-}} in 0, 1, 2 Li-coordination environments as obtained from quantum-chemical calculations.}
\label{fig:1}
\end{center}

The above electronic structure analysis indicates that modulating the LiPFBS concentration in the electrolyte might allow for the adjustment of the distribution of species with specific coordination structures. However, it is important to note that the solvation structures must be quantitatively accounted for since concentration can significantly affect species distribution in an electrolyte system. To this end, we designed computational experiments employing LiPFBS concentrations of 1, 3, and 5~M in DME solvent. Additionally, a 1~M LiTFSI/DME electrolyte was selected for comparative analysis. Classical molecular dynamics simulations were conducted for these systems to investigate solvation structures qualitatively.

\begin{center}
\includegraphics[width=1.08\textwidth]{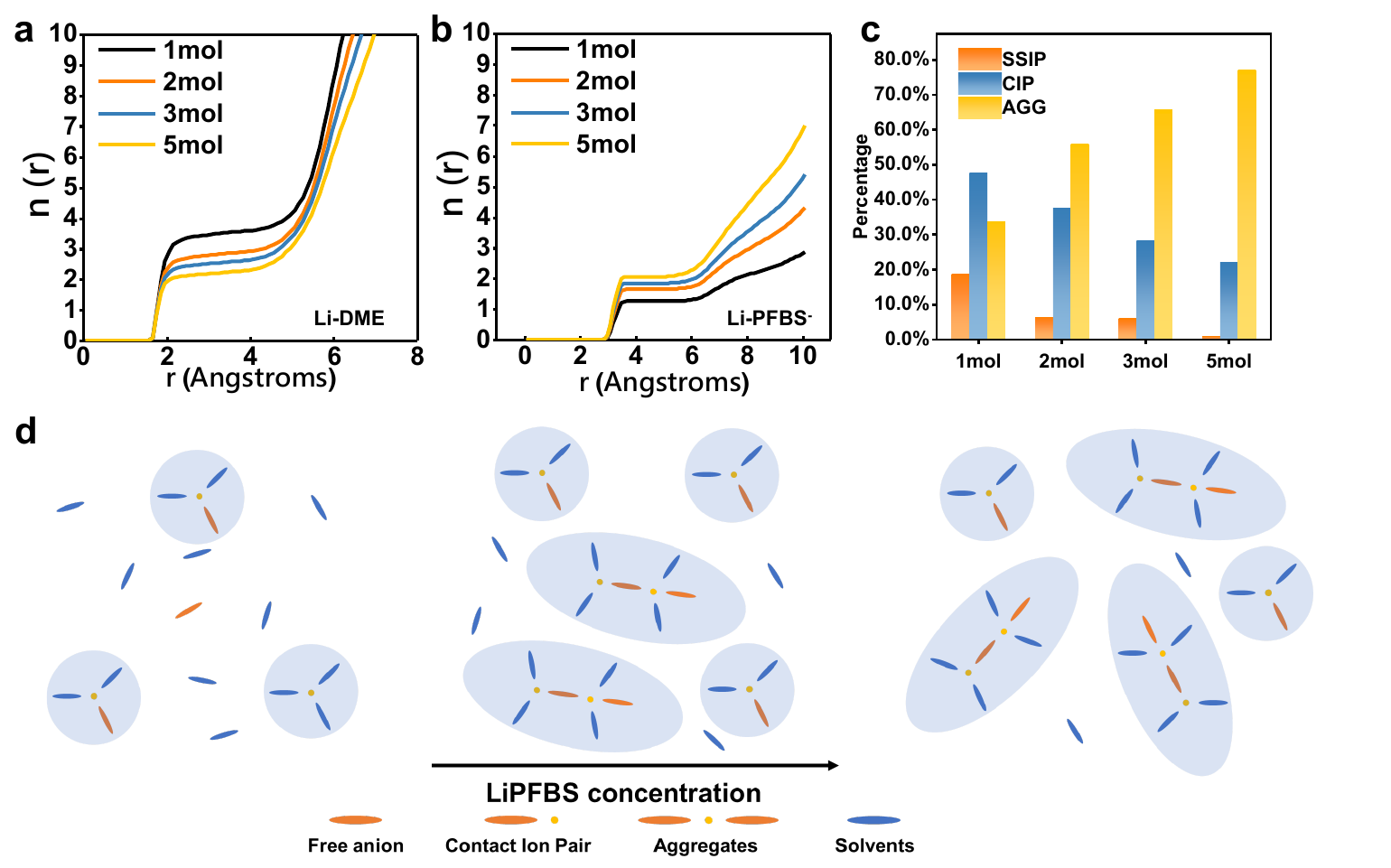}
\captionof{figure}{Li-coordination and solvation structure with varying salt concentration. Li coordination number obtained from the integrated Li-O radial distribution function for \textbf{(a)}~\ce{Li^+}-DME and \textbf{(b)}~\ce{Li^+}-\ce{PFBS^{-}} contact ion pairs at LiPFBS concentrations between 1~and~5~mol. \textbf{(c)}~distribution of \ce{PFBS^{-}} separated ion pair (SSIP), contact ion pair (CIP), and aggregates (AGG). \textbf{(d)}~Schematic of the solvation-structure variations with increasing LiPFBS concentration.}
\label{fig:2}
\end{center}

Initially, classical MD simulations were conducted to investigate the species distribution in DME solvent at varying LiPFBS concentrations of 1, 3, and 5 M, respectively. As seen in \textbf{Figure~\ref{fig:2}}, the integrated radial distribution functions (RDFs) for 1, 3, and 5~M electrolytes reveal that the average coordination number of DME around each \ce{Li^{+}} ion diminished from approximately 3.5 to 2 as the salt concentration increases (\textbf{Figure~\ref{fig:2}a}). Conversely, the Li coordination number for \ce{PFBS^{-}} increased from around 1.3 to 2.2 (\textbf{Figure \ref{fig:2}b}).

We further analyzed the MD trajectories to determine the distribution of separated ion pairs, contact ion pairs (\ce{Li^{+}}-\ce{PFBS^{-}}), and aggregates ($n$\ce{Li^{+}}-$m$\ce{PFBS^{-}}), where $n, m$ $\geq$ 2) within the liquid electrolyte phase. With increasing concentration, the population of free anions diminished while the occurrence of contact ion pairs and aggregates increased.

Together with the electronic structure calculations for 0, 1, and 2-coordinated DME and \ce{PFBS^{-}} presented in \textbf{Figure \ref{fig:1}}, it can be inferred that as concentration increases, the number of coordinated DME molecules decreases, the number of contact ion pairs and aggregates increases, and the propensity for SEI formation shifts towards being predominantly governed by multi-coordinated \ce{PFBS^{-}}.

To validate the computational predictions, we investigated the relationship between LiPFBS concentration and the resulting SEI experimentally. XPS characterization was carried out for the composition of SEI formed by different concentrations of LiPFBS/DME electrolytes and 1~M LiTFSI/DME electrolyte for comparison. As shown in \textbf{Figure~\ref{fig:3}}, the main elemental composition of the SEI formed on the electrode surface of a symmetric Li$||$Li cell after cycling includes Li, S, O, C, and F. Compared to 1~M LiTFSI/DME, the elemental distribution of the resulting SEI from 1, 3 and 5 M LiPFBS/DME is significantly different. The SEI elemental distribution of C and O indicates that 1~M LiTFSI/DME electrolyte has a higher organic component which could be attributed to DME reduction products\cite{renHighconcentrationEtherElectrolytes2019}.

The observed trends are in agreement with the computational predictions shown in \textbf{Figure~\ref{fig:2}c}, since at a \ce{PFBS^{-}} concentration of 1~M, more than 30\% of the anions for aggregates (AGG), which have approximately the same LUMO energy as Li-coordinated DME. \ce{PFBS^{-}} anions tend to participate more in SEI formation as LiPFBS concentration increases. Compared to the 1~M LiPFBS/DME electrolyte, 3~M and 5~M LiPFBS/DME electrolytes indeed formed SEIs with increased F contents and a similar Li content due to changes in anion solvation affecting the reduction potential, making it easier to participate in the formation of the SEI. In addition, as shown in \textbf{Figures~S1-4}, we observed that the F content is less near the electrode surface by comparing different etching depths. This indicates that in the early stage of SEI formation, the reduction of \ce{PFBS^{-}} is less common when electron migration occurs without the formed SEI. After the SEI has been partially formed, the reduction of \ce{PFBS^{-}} becomes more dominant as the hindrance to electron migration increases. In addition, SEI derived from LiPFBS/DME electrolytes exhibited a higher fluorine content than those originating from LiTFSI/DME across varying concentrations. Moreover, this fluorine content increased with the electrolyte concentration, which also can be attributed to more \ce{PFBS^{-}} reduction in an earlier stage of SEI formation. However, as shown in the C1s spectra, it can be observed that DME can still participate in SEI formation by a large percentage due to DME generally lower LUMO energy in various Li coordination environments.

\begin{center}
\includegraphics[width=1.05\textwidth]{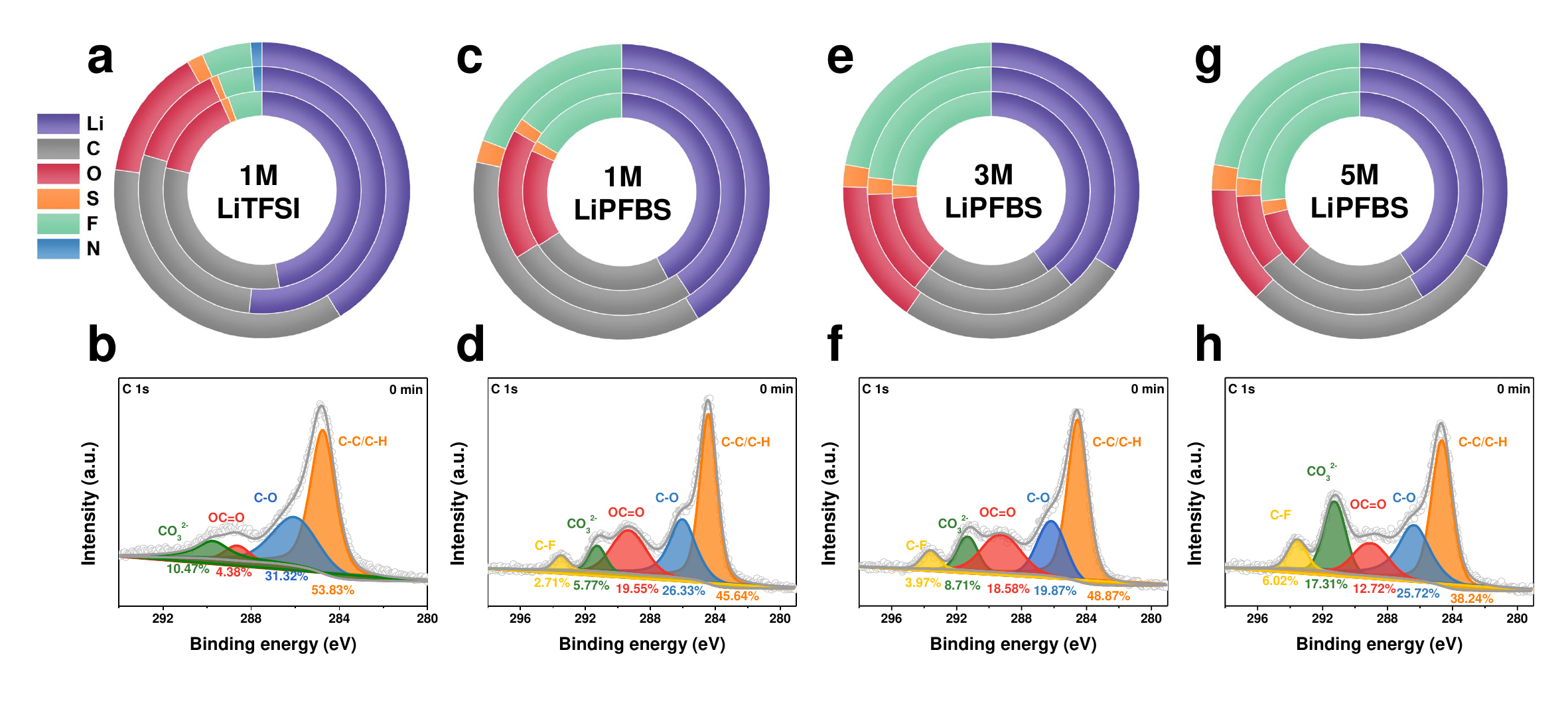}
\captionof{figure}{Elemental distribution of the SEI formed on the Li anode after 10 cycles in symmetric Li$||$Li cells and the corresponding C1s XPS spectra for \textbf{(a-b)}~1~M LiTFSI/DME, \textbf{(c-d)}~1 M LiPFBS/DME, \textbf{(e-f)}~3~M LiPFBS/DME, and \textbf{(g-h)}~5~M LiPFBS/DME.}
\label{fig:3}
\end{center}

Since the interaction between \ce{Li^{+}} and DME or \ce{PFBS^{-}} can notably influence orbital distribution, leading to changes in corresponding electrochemical redox behavior, we have delved deeper into the one-electron reduction mechanism of DME and \ce{PFBS^{-}} within various solvation environments, as illustrated in \textbf{Figure~\ref{fig:4}}. The theoretical reduction potential derived from the proposed one-electron reduction mechanism suggests that, as indicated by the HOMO-LUMO energy trends, the coordination of \ce{Li^{+}} can significantly boost the reductive reactivity of both DME and \ce{PBFS^{-}}. Notably, the typical CIP form of LiPFBS exhibits a higher theoretical reduction potential than the typical AGG form of LiPFBS. This observation may explain why the resulting SEI from an electrolyte with a very high concentration of LiPFBS may not be primarily derived from \ce{PFBS^{-}}, as suggested by the analysis of the Li anode XPS spectra.

\begin{center}
\includegraphics[width=1.0\textwidth]{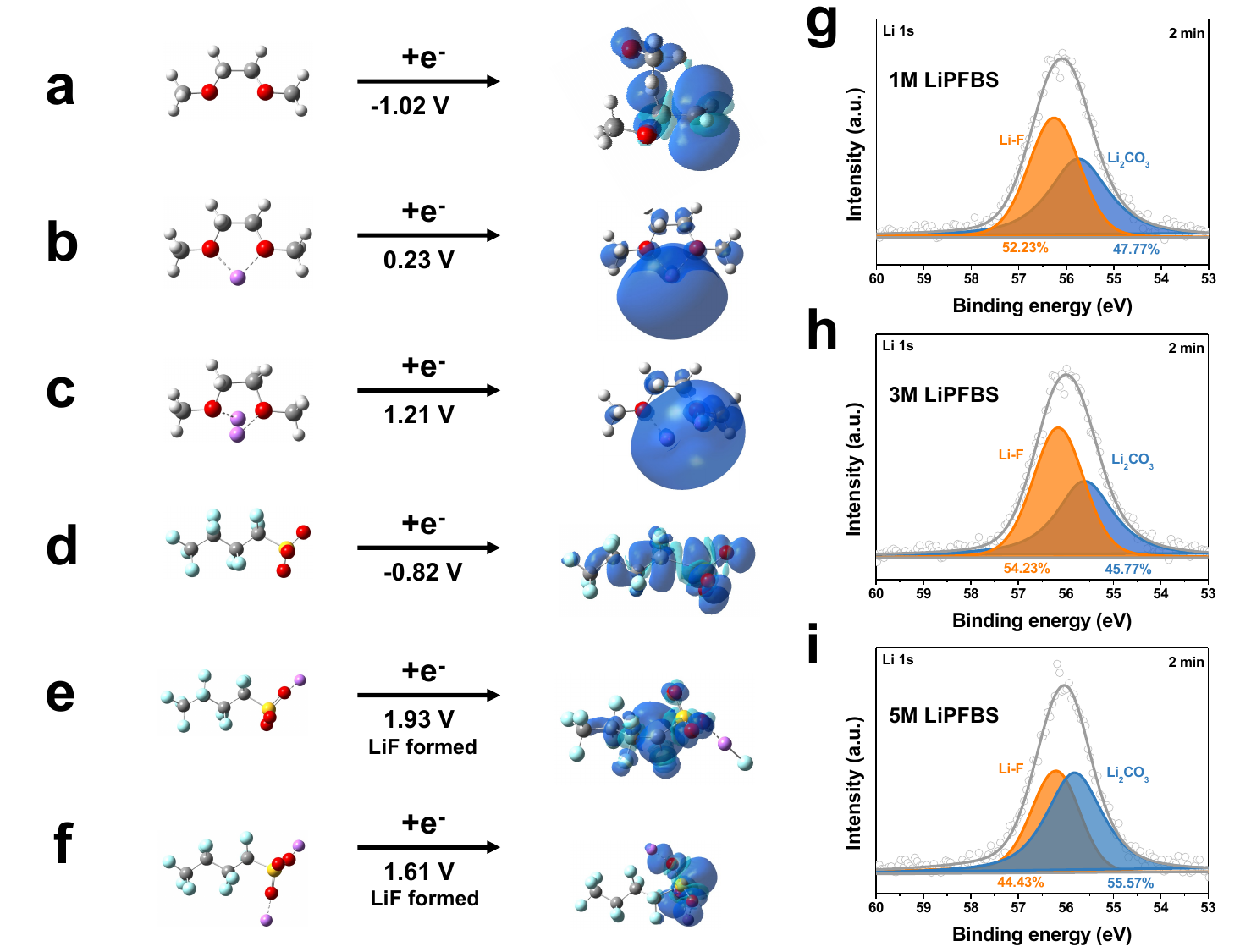}
\captionof{figure}{Proposed one electron reduction mechanisms of \textbf{(a)}~DME \textbf{(b)}~\ce{Li^{+}-DME}, \textbf{(c)}~\ce{2Li^{+}-DME} , \textbf{(d)}~\ce{PFBS^{-}}, \textbf{(e)}~\ce{Li^{+}-PFBS^{-}}, and \textbf{(f)}~\ce{2Li^{+}-PFBS^{-}}. \textbf{(g-i)}~Li 1s XPS spectra of Li anodes for 1~M LiPFBS/DME, 3~M LiPFBS/DME and 5~M LiPFBS/DME after 10 cycles in Li$||$Li symmetric cells.}
\label{fig:4}
\end{center}

\begin{center}
\includegraphics[width=1\textwidth]{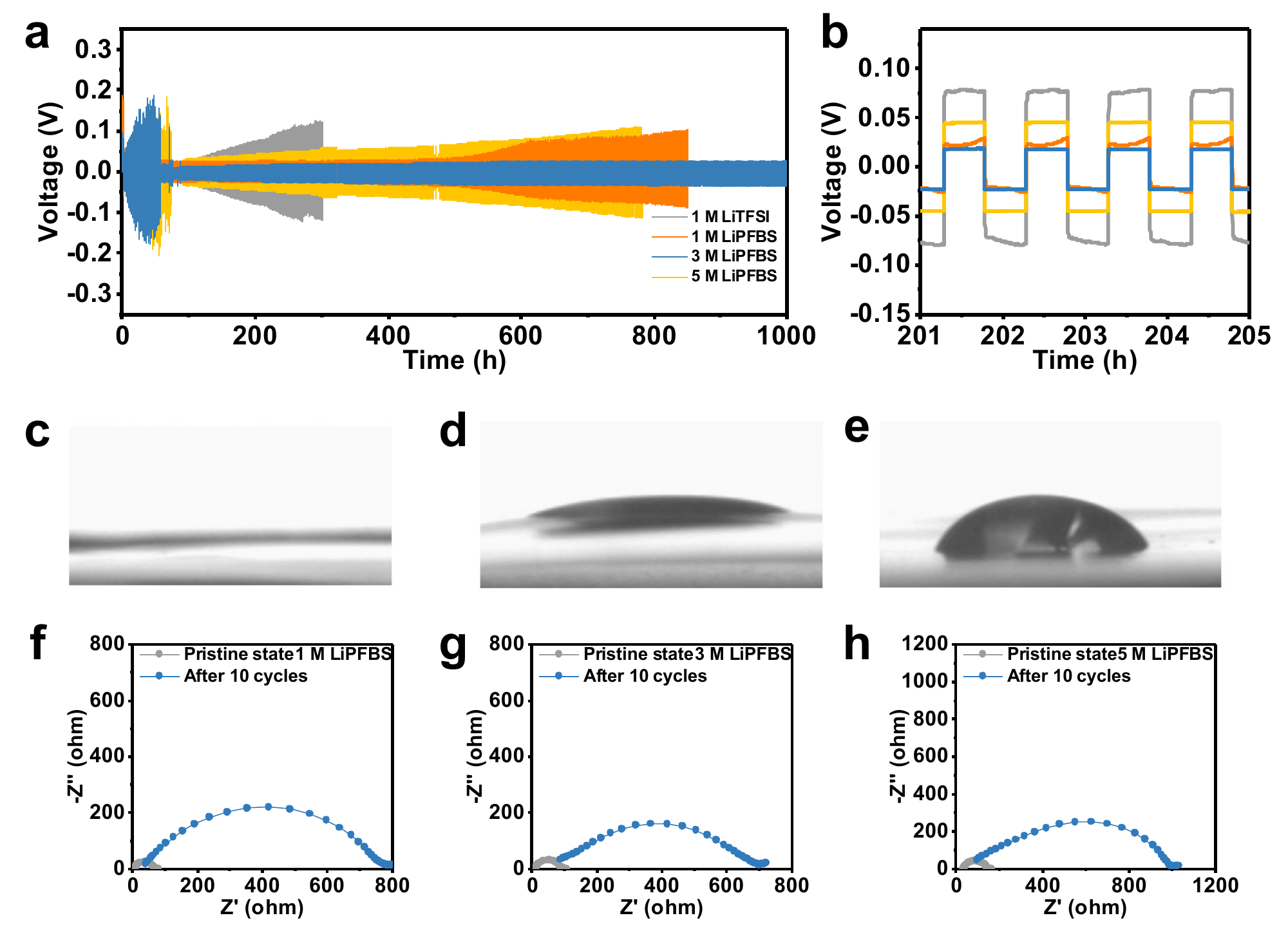}
\captionof{figure}{\textbf{(a)}~Li$||$Li symmetric cell cycling performance and \textbf{(b)}~detailed voltage profiles in different electrolytes, including 1~M LiTFSI, 1~M LiPFBS, 3~M LiPFBS and 5~M LiPFBS at a current density of 0.5~mA\,cm$^{-2}$ with a capacity of 0.5~mAh\,cm$^{-2}$, \textbf{(c-e)}~contact angle of 1, 3, and 5~M LiPFBS/DME electrolyte, \textbf{(f-h)} EIS spectra of 1, 3, and 5~M LiPFBS/DME electrolyte in pristine state and after 10 cycles.} \label{fig:5}
\end{center}

To further investigate the performance of the SEI formed by different concentrations of LiPFBS/DME electrolytes, 1, 3, and 5~M LiPFBS/DME, as well as 1 M LiTFSI/DME, we characterized symmetric Li$||$Li cells with the various electrolytes. As shown in \textbf{Figure \ref{fig:5}}a, the cell with a 1~M LiTFSI/DME only achieved 300 cycles and exhibited a dramatic increase in overpotential. On the other hand, 1~M LiPFBS/DME and 5~M LiPFBS/DME can support 790 and 824 cycles, respectively. The cell with the 3~M LiPFBS/DME electrolyte exhibited the best cycling stability, supporting 1000 cycles with a low polarization. The voltage of the symmetric cell in LiPFBS/DME electrolytes during cycles 201-205 showed significantly smaller polarization than that of the LiTFSI/DME electrolyte (\textbf{Figure \ref{fig:5}b)}. A likely reason for this difference is that LiPFBS electrolytes yield F-rich SEI films that effectively suppress the growth of lithium dendrites (\textbf{Figure S1-S4}), while LiTFSI salt does not. Due to this effect, the growth of lithium dendrites and the formation of dead lithium cannot be suppressed in the LiTFSI electrolyte, leading to a significant increase in the voltage polarization of the battery.

As shown in \textbf{Figures \ref{fig:5} c-e} and Table S1, the contact angle of 1, 3, and 5 M LiPFBS/DME increases as the salt concentration increases, indicating an increase in viscosity and a decrease in conductivity. The overpotential in Li$||$Li cells is not directly correlated to the viscosity and conductivity (Table S1) of bulk electrolytes, and the ionic conductivity and viscosity do not always exhibit a linear correlation\cite{jangHighVoltageCompatibleDualEtherElectrolyte2021}. Instead, the impedance from the SEI can also play a major role. EIS spectra in \textbf{Figures \ref{fig:5}f-h} show the overall impedance before and after SEI formation. The impedance of pristine state LiPFBS/DME electrolytes has a linear correlation with viscosity. However, after 10 cycles, when the SEI has been formed, the 3~M LiPFBS/DME electrolyte exhibits lower impedance in EIS as well as lower overpotential in the half cell. This indicates that the ionic conductivity in 3~M LiPFBS/DME-derived SEI reduces the overall impedance due to better Li transport kinetics in the SEI, which can also be attributed to a more uniform SEI morphology observed in the SEM image (\textbf{Figure S5}).

\begin{center}
\includegraphics[width=0.8\textwidth]{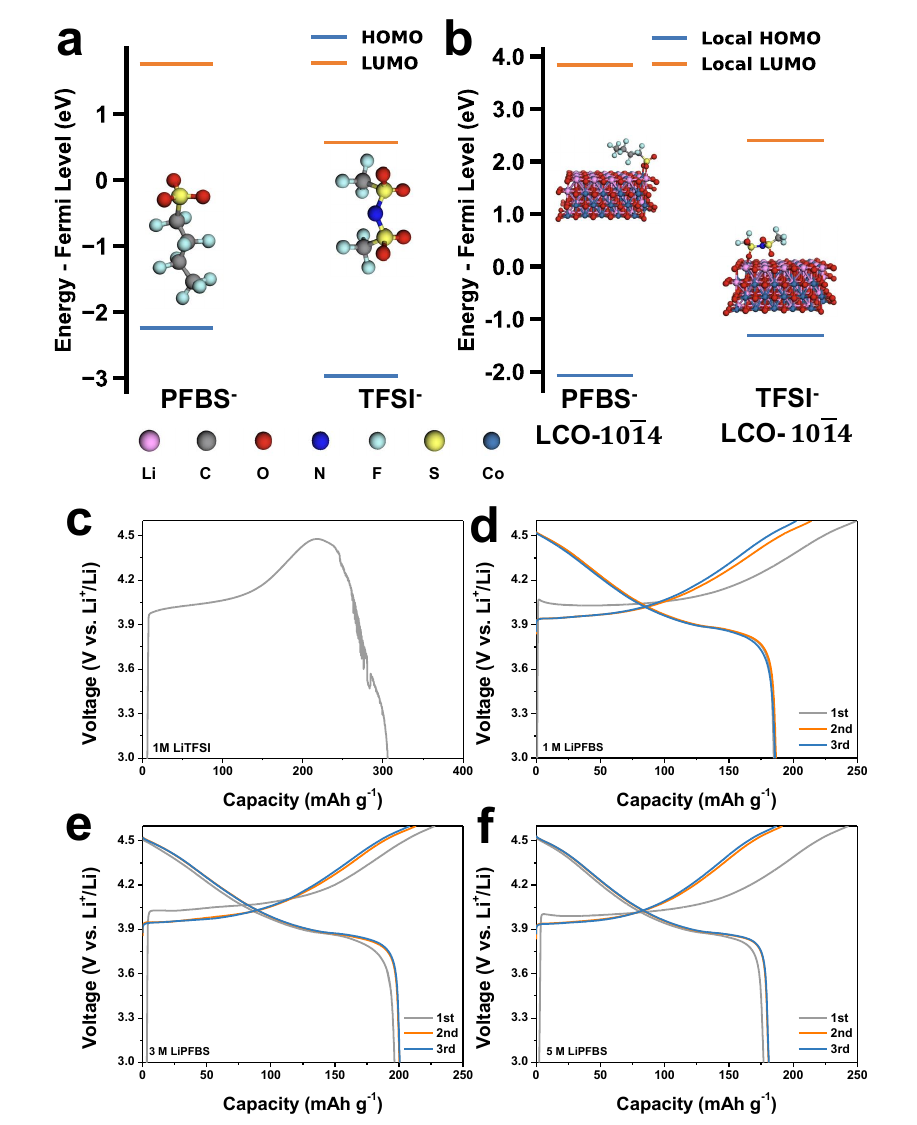}
\captionof{figure}{\textbf{(a)} HOMO and LUMO energies of isolated \ce{PFBS^{-}} and \ce{TFSI^{-}}; \textbf{(b)} local HOMO and LUMO of \ce{PFBS^{-}} and \ce{TFSI^{-}} of the corresponding adsorbed state on the \ce{LiCoO2}($10\ensuremath{\overline{1}}4$) surface; \textbf{(c-f)}~charge-discharge curve of 1~M LiTFSI/DME, 1, 3, and 5~M LiPFBS/DME electrolytes in Li$||$LCO full cells, respectively.}
\label{fig:6}
\end{center}

To further investigate the mechanisms underlying cathode-electrolyte interface (CEI) formation in LiPFBS/DME electrolytes with varying salt concentrations, as well as in 1 M LiTFSI/DME, the HOMO-LUMO energies of \ce{PFBS^{-}} and \ce{TFSI^{-}} in both isolated and adsorbed states were investigated, and the results are shown in \textbf{Figure \ref{fig:6}a-b}. \textbf{Figure \ref{fig:6}c-f} shows measured voltage curves for Li$||$LCO full cells. Unlike SEI formation, CEI formation is dominated by the HOMO energy of the anion, which is concentrated on the cathode surface\cite{ruanSolventAnionChemistry2023}. \ce{PFBS^{-}} has a higher HOMO energy in the isolated state than \ce{TFSI^{-}}, however \ce{TFSI^{-}} has a higher local HOMO state in the adsorbed state. When the cut-off voltage was set between 3.0 and 4.6~V, the 1~M LiTFSI/DME electrolyte exhibited continuous oxidation during the initial cycle. This suggests that the CEI derived from 1~M LiTFSI/DME cannot withstand the volumetric changes of the \ce{LiCoO2} electrode during cycling, compromising its passivating capabilities. In contrast, LiPFBS/DME electrolytes at concentrations of 1, 3, and 5~M demonstrated stable charge-discharge behavior, indicative of a robust CEI capable of accommodating significant volumetric changes at the cathode under high-voltage conditions. The CEI derived from LiPFBS/DME was formed at a lower voltage and is more inorganic, as discussed below.

\begin{center}
\includegraphics[width=1\textwidth]{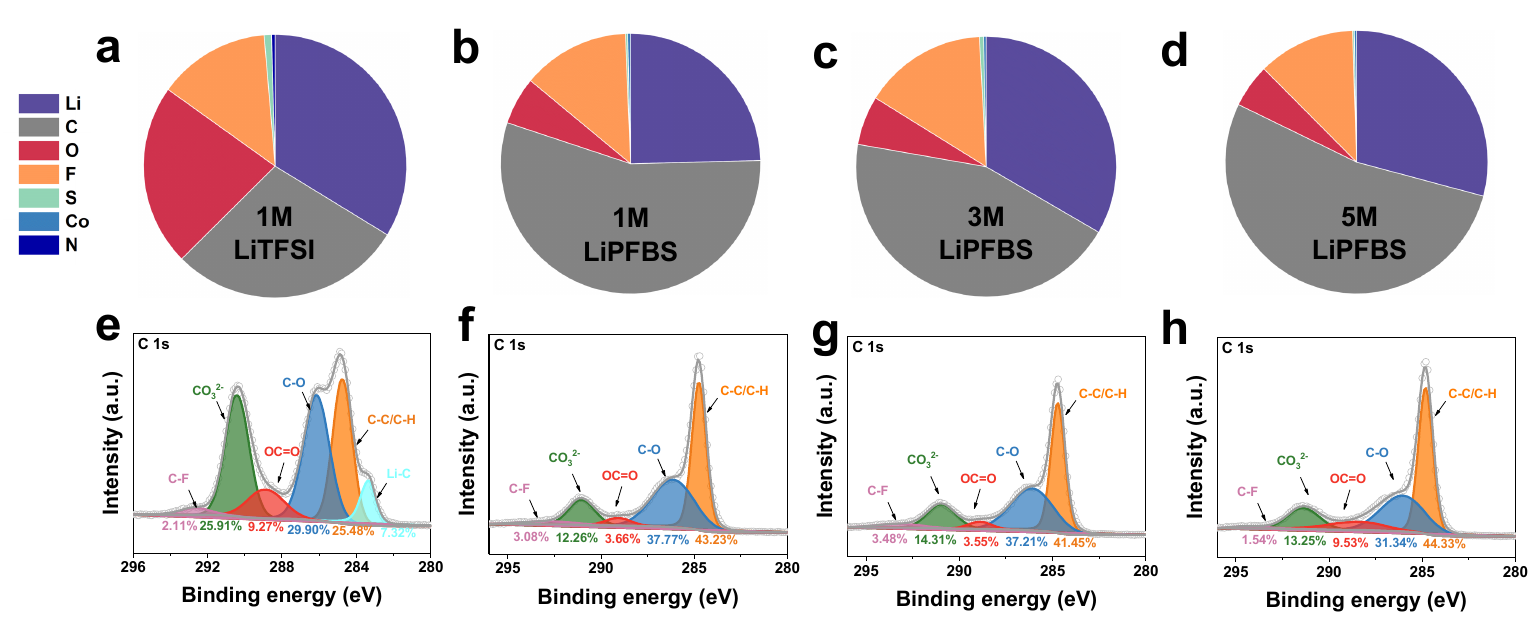}
\captionof{figure}{\textbf{(a)}~Elemental distribution of the CEI on the \ce{LiCoO2} electrode in 1~M LiTFSI/DME electrolyte after the first cycle. \textbf{(b-d)}~Elemental distribution of the CEI in 1, 3, and 5~M LiPFBS/DME after 10 cycles, respectively. \textbf{(e-h)}~Corresponding C 1s XPS spectra from measurements of the \ce{LiCoO2} surface in 1~M LiTFSI/DME, 1~M LiPFBS/DME, 3~M LiPFBS/DME, and 5~M LiPFBS/DME, respectively.}
\label{fig:7}
\end{center}

The stability of CEI is a decisive factor for cell performance in high cut-off voltage in Li metal batteries. It is believed that DME cannot form a passivating CEI layer to prevent continuous electrolyte oxidation\cite{amanchukwuNewClassIonically2020}. Therefore, it is essential to establish a correlation between CEI composition and stability. As shown in \textbf{Figure \ref{fig:7}}, compared to LiPFBS/DME electrolyte, LiTFSI yielded a CEI with significantly higher O content. Meanwhile, the C~1s spectra indicate a high ratio of DME oxidation products, which correspond to the \ce{CO_{3}^{2-}}, \ce{OC=O}, and \ce{C-O} peaks. These three peaks indicate an organic-rich CEI, which causes continuous electrolyte oxidation under 3.0-4.6~V charging. In contrast, \textbf{Figures~\ref{fig:7}b-d and 7f-h} show that LiPFBS/DME electrolytes with different salt concentrations yielded CEIs that have lower O contents and form fewer DME oxidation products. Instead, inorganic-rich and stable CEIs were observed, which remained stable in the 2nd and 3rd cycles. Additionally, as shown in \textbf{Figure~S7}, \ce{PFBS^{-}} exhibits a lower density of states for oxygen near the Fermi level, whereas \ce{TFSI^{-}} shows overlapping densities of states for oxygen and nitrogen. However, the elemental percentage of nitrogen is nearly undetectable in the XPS spectra of \ce{LiCoO2} in a 1~M LiTFSI/DME electrolyte. This suggests that the 1~M LiTFSI electrolyte forms a soluble oxidation product that cannot act as a passivation layer to prevent further continuous oxidation.

\begin{center}
\includegraphics[width=1\textwidth]{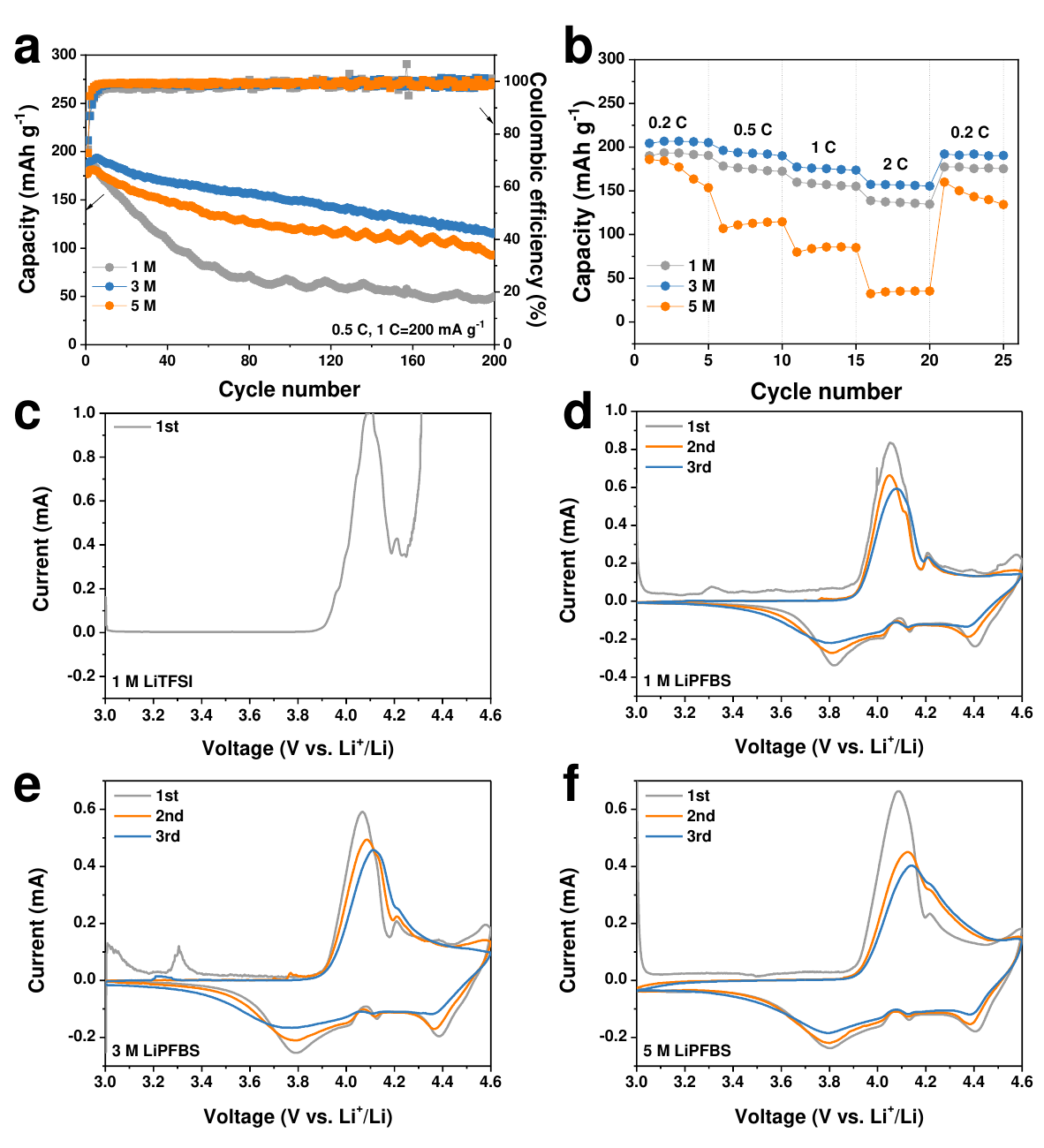}
\captionof{figure}{\textbf{(a, b)}~Cycling performance and rate performance of Li$||$LCO full cells with 1, 3, 5~M LiPFBS/DME electrolytes. Cyclic voltammetry curves of Li$||$LCO in \textbf{(c)}~1~M LiTFSI/DME, \textbf{(d)}~1~M LiPFBS/DME, \textbf{(e)}~3~M LiPBFS/DME, \textbf{(f)}~5~M LiPBFS/DME electrolyte. }
\label{fig:8}
\end{center}

Even though the cathode itself is also a factor that can cause battery failure, we chose \ce{LiCoO2} as the cathode to examine CEI performance because \ce{LiCoO2} has decent high-voltage stability compared to \ce{LiNi_{x}Mn_{y}Co_{z}O2} (NMC) materials. As shown in \textbf{Figure~\ref{fig:8}c-f}, LiTFSI/DME exhibits the same oxidation behavior in a good agreement with Wang et.al\cite{wang_high-voltage_2022}, the main oxidation peak at 4.1V and a smaller peak after at approximately 4.2 V. While LiPFBS/DME electrolytes show a stable charge-discharge curve and a lower oxidation peak slightly below 4.1 V, the second small peak remains at approximately 4.2 V. The main oxidation peak shift can be attributed to the earlier oxidation of \ce{PFBS^{-}} because the phase transition of \ce{LiCoO2} usually occur at above 4.5 V\cite{xiaPhaseTransitionsHighVoltage2007}. Despite DFT calculations showing that \ce{PFBS^{-}} has a lower local HOMO on the \ce{LiCoO2}($10\ensuremath{\overline{1}}4$) surface than \ce{TFSI^{-}}, it is important to note that the shift in oxidation potential cannot be solely attributed to anion oxidation, though it is a crucial factor. As we discussed previously, an inorganic-rich protective CEI layer derived by \ce{PFBS^{-}} prevents DME from continuous oxidation and stabilizes \ce{LiCoO2} at high voltage. However, the performance of the Li$||$LCO full cell is also determined by SEI properties. Therefore, it makes sense that the 3~M LiPFBS/DME has the best capacity retention, Coulombic efficiency, and rate performance, which can be attributed to lower impedance owed to a highly conductive SEI. Hence, it is obvious that the rate performance of LiPFBS/DME is linearly correlated to the overall impedance from the bulk electrolyte and SEI; in contrast, capacity retention correlates more with CEI stability and cathode degradation. As shown in \textbf{Figure \ref{fig:8}a-b}, 3~M LiPFBS exhibits the best capacity retention rate of 73$\%$ after 200 cycles, as well as the best rate capability among 1, 3, 5 M LiPFBS/DME electrolyte.

\section*{Conclusion}

In summary, we have comprehensively investigated a novel LiPFBS/DME electrolyte system at different salt concentrations, including its SEI/CEI formation mechanism, electrochemical behavior, and cell performance. We have engineered a high-concentration electrolyte capable of enabling 4.6~V Li$||$LCO batteries. By integrating DFT calculations, MD simulations, and X-ray photoelectron spectroscopy (XPS), we provided a thorough perspective on SEI/CEI formation and its correlation with cell performance. Compared to 1~M LiTFSI/DME, LiPFBS/DME exhibits a lower HOMO energy level and preferentially oxidizes at the cathode surface, making the high-voltage application of DME realistic. Additionally, the benefits of a stable DME-derived SEI are still preserved, providing decent full-cell performance under a 3.0-4.6~V working voltage. We anticipate that this work can guide general electrolyte design and pave the path for next-generation high-energy-density Li metal batteries.

\section*{Data Availability Statement}

The authors declare no competing financial interest.
The data from the Density Functional Theory (DFT) calculations can be obtained from the GitHub repository at \\ https://github.com/atomisticnet/2024-LiCoO-electrolytes-Li-DFT-Data/. The dataset contains atomic structures in VASP and PDB format.

\section*{Acknowledgement}
J.H. and N.A. thank the Dutch National e-Infrastructure and the SURF Cooperative for the computational resources used in the DFT calculations. All experimental results were obtained by S.Q. and J.M., supported by the National Natural Science Foundation of China (Grant No. U21A20311). This work was funded by a start-up grant (Dutch Sector Plan) from Utrecht University awarded to N.A.

\clearpage\newpage
\appendix
\renewcommand{\thesection}{S\arabic{section}}
\renewcommand{\thesubsection}{S\arabic{section}}
\renewcommand{\thefigure}{S\arabic{figure}}
\renewcommand{\thetable}{S\arabic{table}}
\setcounter{figure}{0}
\setcounter{table}{0}

\section{Supporting Information}

\begin{table}[h]
    \centering
    \caption{Conductivity of Different Electrolytes: LiTSFI/ dimethoxyethane (DME) and perfluorobutane sulfonate (LiPFBS)/DME electrolytes.}
    \begin{tabular}{|c|c|}
        \hline
        Electrolytes & Conductivity (us/cm) \\
        \hline
        1M LiTFSI/DME & 4880 \\
        1M LiPFBS/DME & 3487 \\
        3M LiPFBS/DME & 1865 \\
        5M LiPFBS/DME & 942 \\
        \hline
    \end{tabular}
\end{table}

\begin{center}
\includegraphics[width=1.1\textwidth]{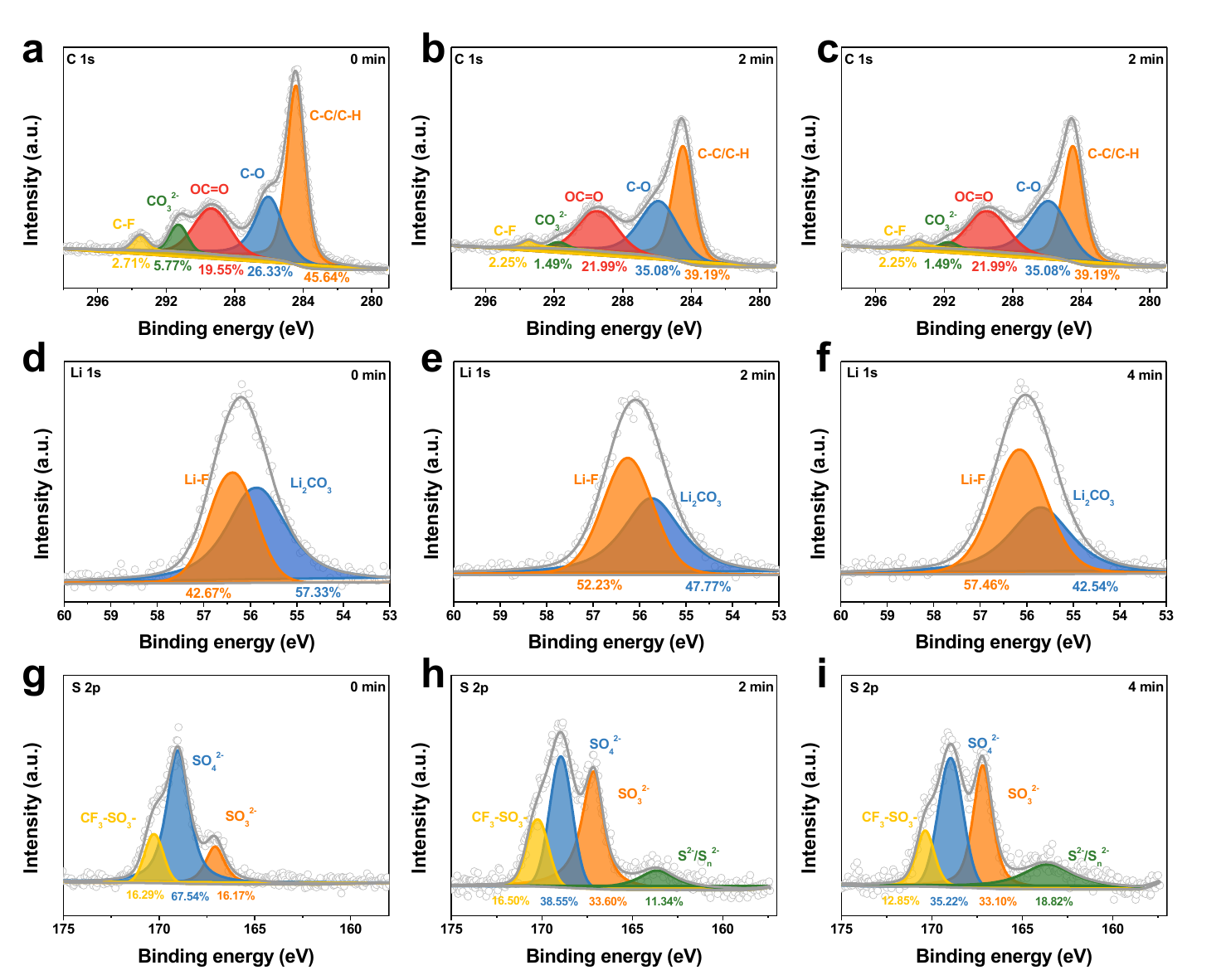}
\captionof{figure}{The XPS spectra of C1s (a-c), Li 1s (d-f), and S 2p (g-i) spectra of the Li anode in \textbf{1M LiTFSI/DME} electrolyte after 10 cycles in Li$||$Li symmetric cells. The Li anode was sputtered with Ar for 0, 2, and 4 minutes to study different depths of the SEI.}
\label{fig:5}
\end{center}

\begin{center}
\includegraphics[width=1.1\textwidth]{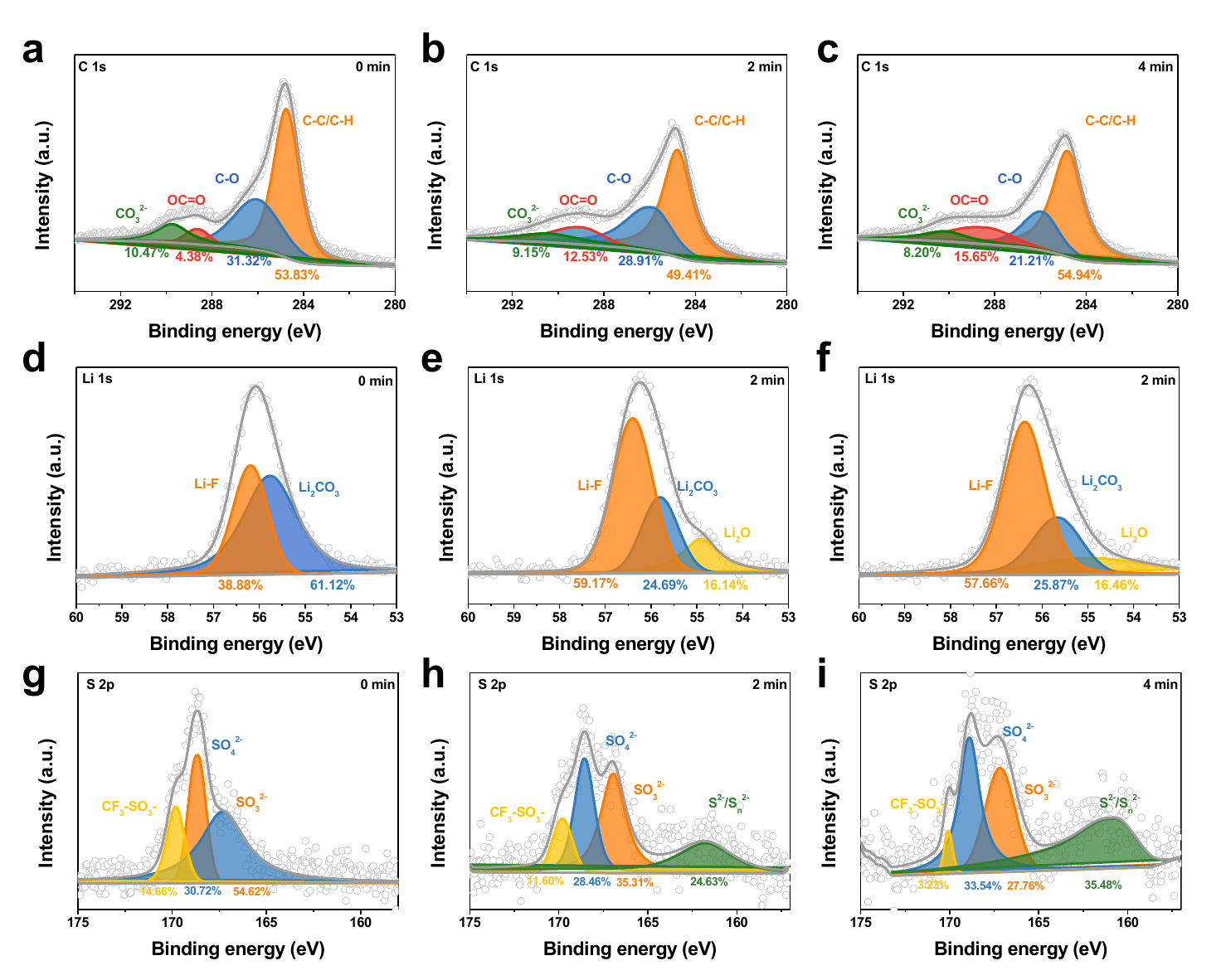}
\captionof{figure}{The XPS spectra of C1s (a-c), Li 1s (d-f), and S 2p (g-i) spectra of the Li anode in \textbf{1M LiPFBS/DME} electrolyte after 10 cycles in Li$||$Li symmetric cells. The Li anode was sputtered with Ar for 0, 2, and 4 minutes to study different depths of the SEI.}
\label{fig:5}
\end{center}

\begin{center}
\includegraphics[width=1.1\textwidth]{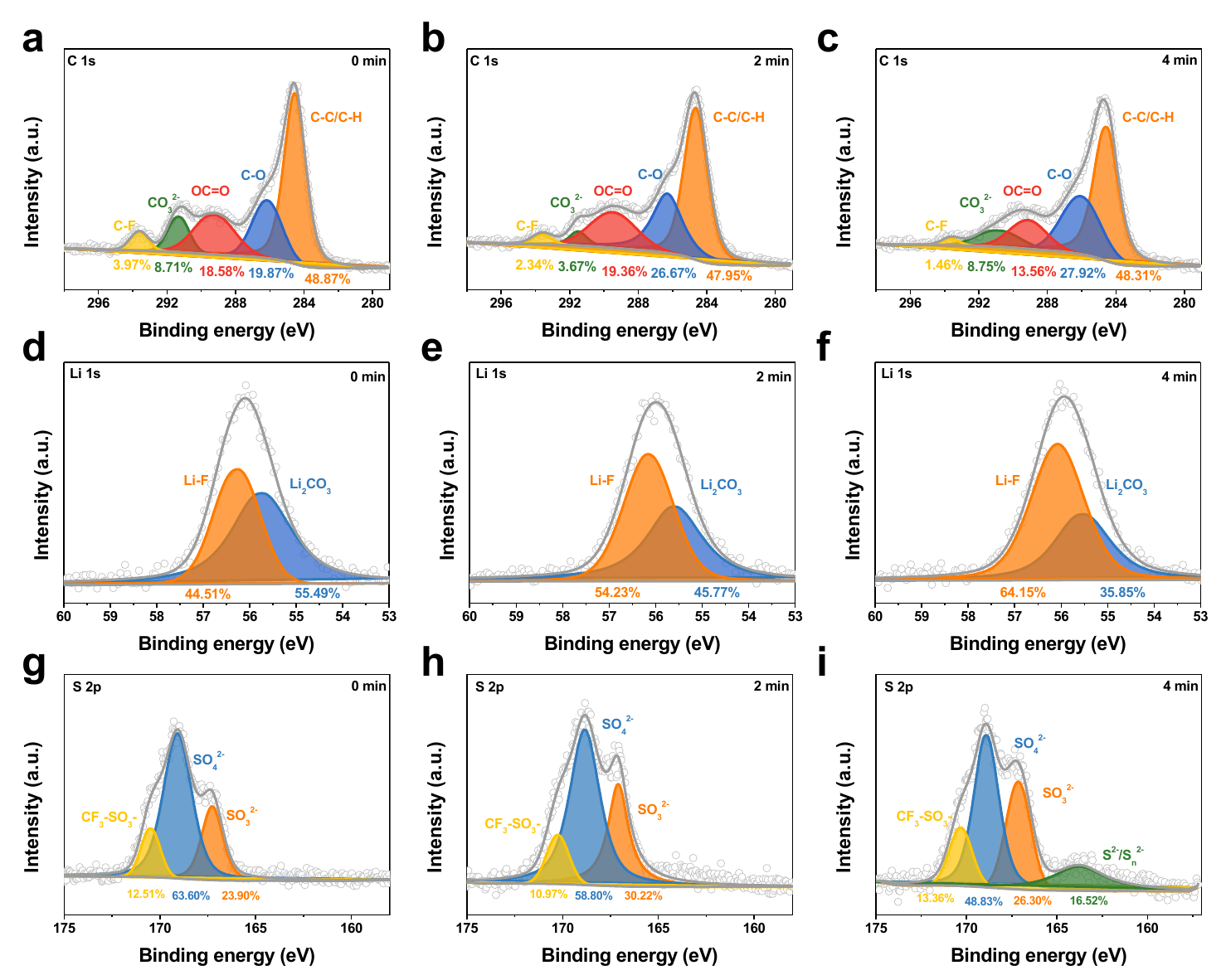}
\captionof{figure}{The XPS spectra of C1s (a-c), Li 1s (d-f), and S 2p (g-i) spectra of the Li anode in \textbf{3M LiPFBS/DME} electrolyte after 10 cycles in Li$||$Li symmetric cells. The Li anode was sputtered with Ar for 0, 2, and 4 minutes to study different depths of the SEI.}
\label{fig:5}
\end{center}

\begin{center}
\includegraphics[width=1.1\textwidth]{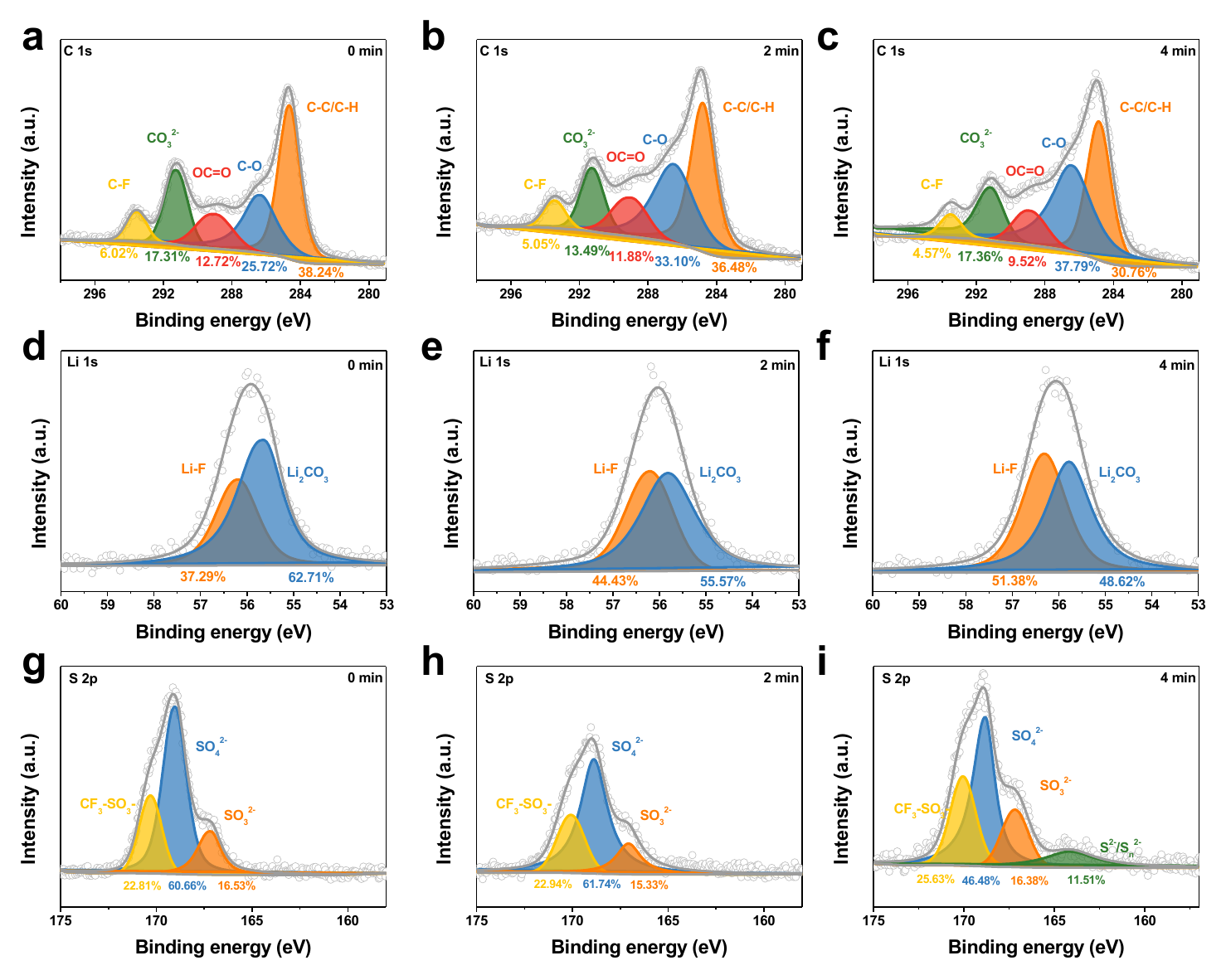}
\captionof{figure}{The XPS spectra of C1s \textbf{(a-c)}, Li 1s \textbf{(d-f)}, and S 2p \textbf{(g-i)} spectra of the Li anode in \textbf{5M LiTFS/DME} electrolyte after 10 cycles in Li$||$Li symmetric cells. The Li anode was sputtered with Ar for 0, 2, and 4 minutes to study different depths of the SEI.}
\label{fig:5}
\end{center}

\begin{center}
\includegraphics[width=0.8\textwidth]{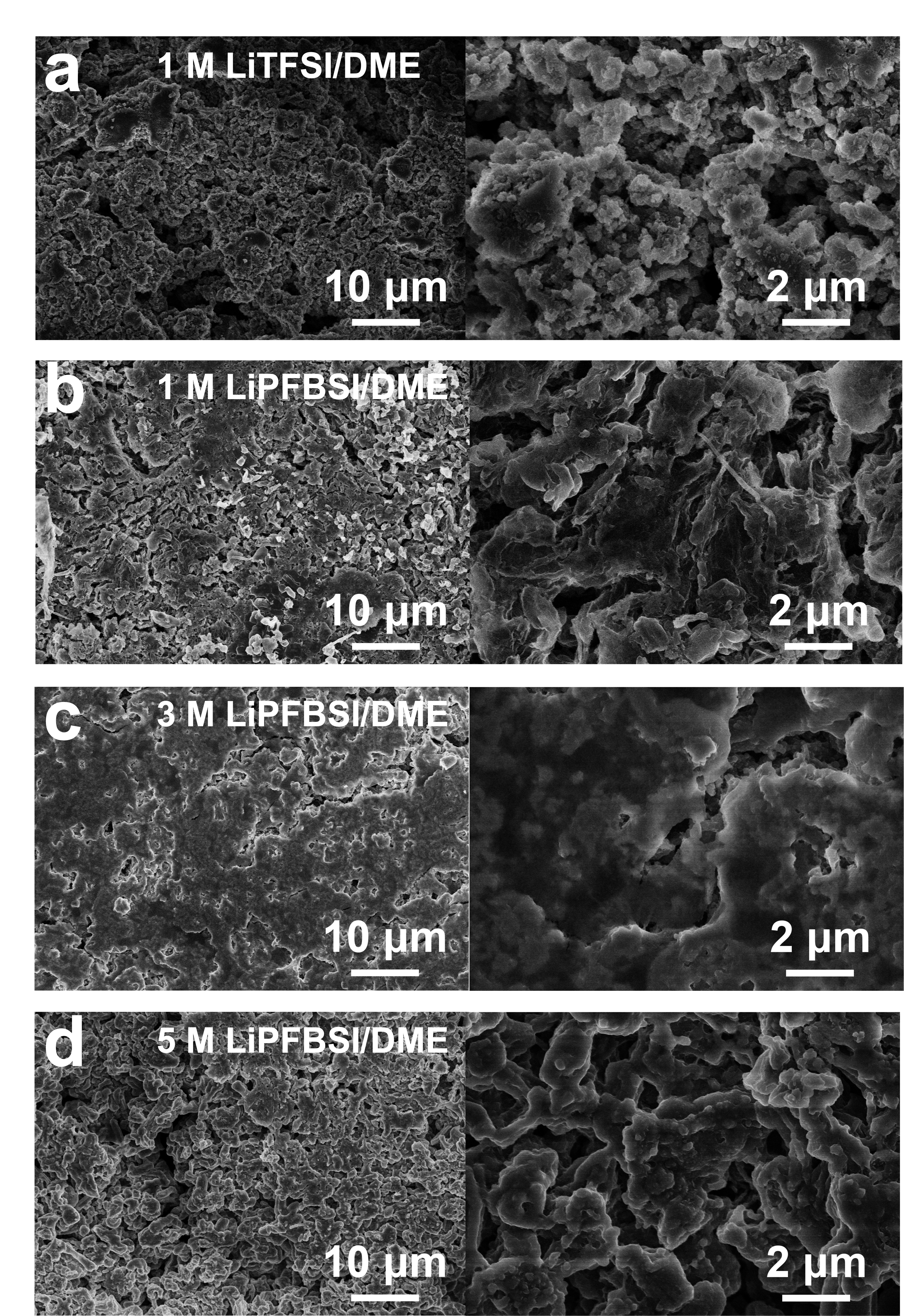}
\captionof{figure}{SEM images of the Li anode in \textbf{5M LiTFSI/DME} electrolyte after 10 cycles in Li$||$Li symmetric cells of \textbf{(a)} 1M LiTFSI/DME, \textbf{(b)} 1M LiPFBS/DME, \textbf{(c)} 3M LiPFBS/DME \textbf{(d)} 5M LiPFBS/DME.}
\label{fig:5}
\end{center}

\begin{center}
\includegraphics[width=0.8\textwidth]{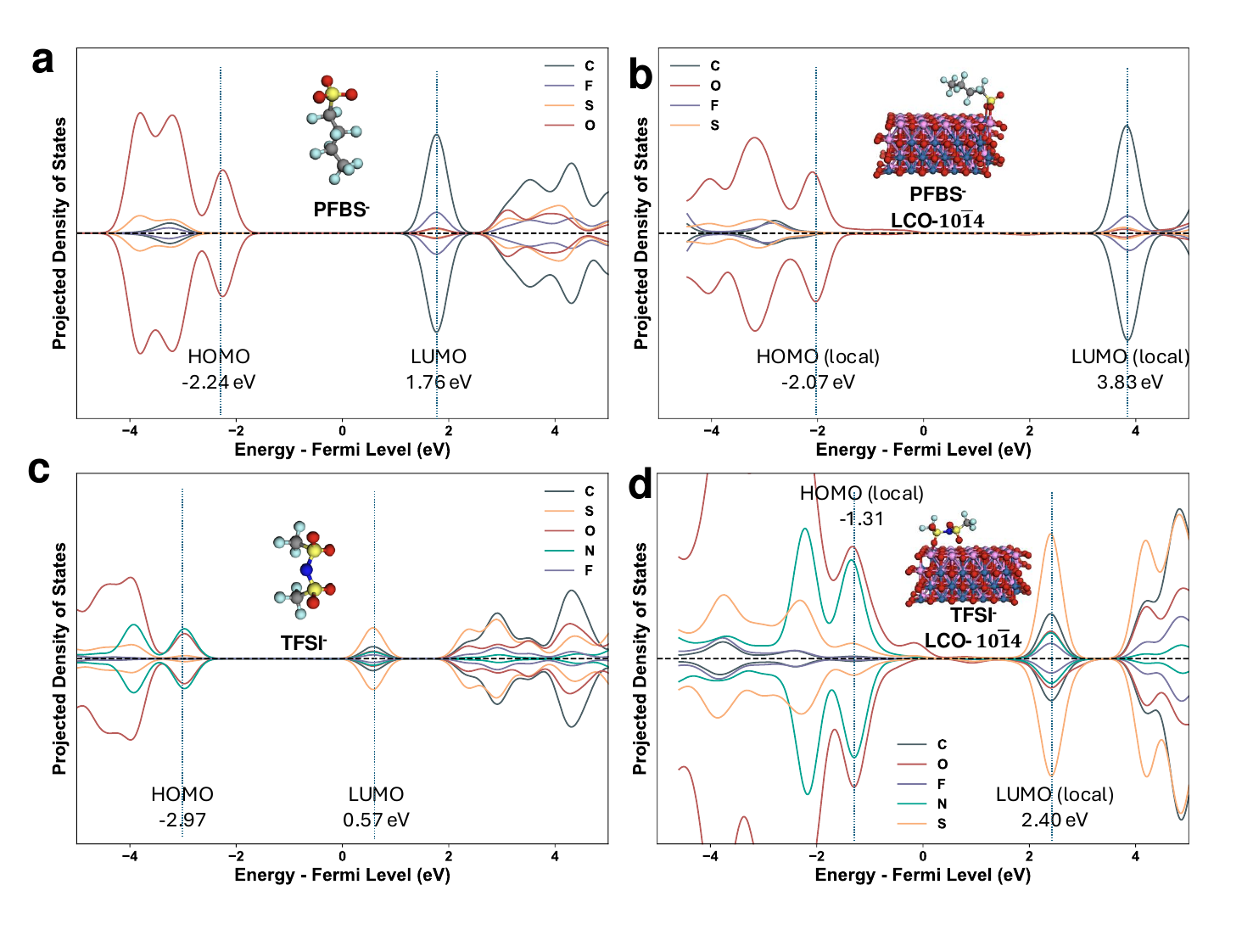}
\captionof{figure}{Projected density of states of isolated \textbf{(a)} \ce{PFBS^{-}} \textbf{(b)} isolated \ce{TFSI^{-}}, \textbf{(c)} \ce{PFBS^{-}} and \textbf{(d)} \ce{TFSI^{-}} on \ce{LiCoO2}-$10\ensuremath{\overline{1}}4$ surface}
\label{fig:5}
\end{center}


\providecommand{\latin}[1]{#1}
\makeatletter
\providecommand{\doi}
  {\begingroup\let\do\@makeother\dospecials
  \catcode`\{=1 \catcode`\}=2 \doi@aux}
\providecommand{\doi@aux}[1]{\endgroup\texttt{#1}}
\makeatother
\providecommand*\mcitethebibliography{\thebibliography}
\csname @ifundefined\endcsname{endmcitethebibliography}
  {\let\endmcitethebibliography\endthebibliography}{}

\end{document}